\documentclass[notitlepage,nofootinbib,preprintnumbers,amssymb,superscriptaddress,eqsecnum]{revtex4-1}
\usepackage{amsfonts,amssymb,amsmath,graphicx,color,bm,ulem}
\definecolor{ultramarine}{rgb}{0.07, 0.04, 0.56}
\definecolor{cadmiumgreen}{rgb}{0.0, 0.42, 0.24}
\definecolor{indigo(dye)}{rgb}{0.0, 0.25, 0.42}
\usepackage[linktocpage=true]{hyperref}
\hypersetup{
colorlinks=true,
citecolor=ultramarine,
linkcolor=cadmiumgreen,
urlcolor=indigo(dye),
}

\newcommand{\f}[2]{\frac{#1}{#2}}  
\newcommand{\mk}[1]{\left( #1 \right)}  
\newcommand{\kk}[1]{\left[ #1 \right]}  
  
\newcommand{\be}{\begin{equation}}  
\newcommand{\ee}{\end{equation}}
\newcommand{\bem}{\begin{pmatrix}}
\newcommand{\eem}{\end{pmatrix}}

\newcommand{\berep}{b}
\newcommand{\garep}{d}

\renewcommand{\O}{\mathcal{O}}
\renewcommand{\L}{\mathcal{L}}
\newcommand{\pa}{\partial}
\newcommand{\Mpl}{M_{\rm Pl}}

\begin{document}

\title{
Weakly-coupled stealth solution in scordatura degenerate theory
}

\author{Hayato Motohashi}
\affiliation{Center for Gravitational Physics, Yukawa Institute for Theoretical Physics, Kyoto University, Kyoto 606-8502, Japan}

\author{Shinji Mukohyama}
\affiliation{Center for Gravitational Physics, Yukawa Institute for Theoretical Physics, Kyoto University, Kyoto 606-8502, Japan}
\affiliation{Kavli Institute for the Physics and Mathematics of the Universe (WPI), The University of Tokyo Institutes for Advanced Study, The University of Tokyo, Kashiwa, Chiba 277-8583, Japan}

\begin{abstract}
In scalar-tensor theories we revisit the issue of strong coupling of perturbations around stealth solutions, i.e.\ backgrounds with the same forms of the metric as in General Relativity but with non-trivial configurations of the scalar field. The simplest among them is a stealth Minkowski (or de Sitter) solution with a constant, timelike derivative of the scalar field, i.e.\ ghost condensation. In the decoupling limit the effective field theory (EFT) describing perturbations around the stealth Minkowski (or de Sitter) solution shows the universal dispersion relation of the form $\omega^2 = \alpha k^4/M^2$, where $M$ is a mass scale characterizing the background scalar field and $\alpha$ is a dimensionless constant. Provided that $\alpha$ is positive and of order unity, a simple scaling argument shows that the EFT is weakly coupled all the way up to $M$. On the other hand, if the structure of the underlining theory forces the perturbations to follow second-order equations of motion then $\alpha=0$ and the dispersion relation loses dependence on the spatial momentum. This not only explains the origin of the strong coupling problem that was recently pointed out in a class of degenerate theories but also provides a hint for a possible solution of the problem. We then argue that a controlled detuning of the degeneracy condition, which we call scordatura, renders the perturbations weakly coupled without changing the properties of the stealth solutions of degenerate theories at astrophysical scales. 
\end{abstract}

\preprint{YITP-19-98, IPMU19-0156}

\maketitle  


\section{Introduction}
\label{sec:intro}

Scalar-tensor theories serve a simple framework of modification of gravity for models of primordial/present accelerating expansion of the Universe, as well as testing gravity at dynamical system in the strong field regime with observations of gravitational waves. 
In the last decade, a lot of efforts has been made to clarify how much general derivative couplings between scalar field and gravity are allowed without pathology. 
One of the central issues is that nondegenerate higher-derivative theories in general suffer from unbounded Hamiltonian, known as the Ostrogradsky theorem~\cite{Ostrogradsky:1850fid,Woodard:2015zca}.
While imposing a degeneracy of the Lagrangian with respect to the highest-order derivatives avoids the assumption of the Ostrogradsky theorem, there still exist ghost degrees of freedom associated with non-highest but higher-order derivatives~\cite{Motohashi:2014opa}.
A certain set of conditions should be imposed on Lagrangian to eliminate all the Ostrogradsky ghost, which is known as the degeneracy condition~\cite{Motohashi:2014opa,Langlois:2015cwa,Motohashi:2016ftl,Motohashi:2017eya,Motohashi:2018pxg}. 
Being built upon the degeneracy condition, degenerate higher-order scalar-tensor (DHOST) theories involve second-order derivatives of the scalar field up to quadratic order~\cite{Langlois:2015cwa} and cubic order~\cite{BenAchour:2016fzp} and evade the Ostrogradsky ghost.

The degeneracy condition is not related to a symmetry in general and thus is expected to be spoiled by quantum corrections, leading to apparent ghost degrees of freedom whose masses are decreasing functions of the amount of deviation from the degeneracy condition. From the effective field theory (EFT) point of view~\cite{Appelquist:1974tg,Weinberg:1978kz,Georgi:1985kw,Polchinski:1992ed}, however, we should restrict our consideration to sufficiently low energies, momenta and amplitudes of fluctuations below a cutoff. Therefore, if there is an apparent ghost degree of freedom and if it has a mass larger than the cutoff of the EFT then it should not be considered as a problem since the properties and even the existence of the apparent ghost are UV sensitive and can be completely altered by the infinite series of higher dimensional operators that become prominent above the cutoff. This  means that, assuming the existence of a good UV completion, one can safely relax the degeneracy condition as far as the deviation from the degeneracy is small enough so that the apparent ghost is heavier than the cutoff scale. In this case, the limit of taking the mass of the apparent ghost infinity corresponds to the standard degenerate theories in which the apparent ghost is eliminated by a set of constraints.

One of interesting properties of higher-derivative scalar-tensor theories is the existence of stealth solutions, which consist of the same forms of the metric as in General Relativity (GR) solutions but with nontrivial scalar field profiles that do not contribute to the stress energy tensor. 
With a trivial or constant scalar field profile, it is possible to derive systematically a set of sufficient conditions for a wide class of arbitrarily higher-order derivative theories to allow the metric same as in GR as an exact solution~\cite{Motohashi:2018wdq}.
Similarly, for stealth solution with nontrivial profile of the scalar field, one can also perform a systematic analysis to identify a subclass of theories that allow a GR solution in a class of theories of interest.
In particular, shift-symmetric theories allow linearly time dependent scalar field profile compatible with static or stationary metric ansatz.
For instance, various stealth black hole solutions were found for shift symmetric Horndeski theory with linearly time dependent scalar field with $X\equiv g^{\mu\nu}\pa_\mu\phi\pa_\nu\phi={\rm const}$.~\cite{Babichev:2013cya,Kobayashi:2014eva}~\footnote{In the context of k-essence, a stealth Schwarzschild solution was found in \cite{Mukohyama:2005rw} in the k-essence limit of the ghost condensate, i.e.\ in the limit $\alpha\to 0$.}
Another class of stealth solutions with the scalar field profile $\phi=\phi(r)$ and $X\ne{\rm const.}$ were also found for non-shift-symmetric Horndeski theory~\cite{Minamitsuji:2018vuw}.
Stealth black holes in DHOST theories have been attracting much attention recently~\cite{BenAchour:2018dap,Motohashi:2019sen,Charmousis:2019vnf,Minamitsuji:2019shy,Takahashi:2019oxz}. Stealth solutions may also be used as seed solutions in a certain generating-solution method based on disformal transformations~\cite{BenAchour:2019fdf}.

Stability analysis of stealth solutions have also been extensively investigated. In the case of stealth solution with $\phi=\phi(r)$ and $X\ne{\rm const.}$ in non-shift-symmetric Horndeski theory found in \cite{Minamitsuji:2018vuw}, since the scalar field profile $\phi=\phi(r)$ is static, one can apply general results of odd- and even-parity perturbation theory around static, spherically symmetric spacetime in Horndeski theory established in~\cite{Kobayashi:2012kh,Kobayashi:2014wsa}. 
It was found in \cite{Minamitsuji:2018vuw} that the kinetic term of the second even mode vanishes, signaling strong coupling.
The strong coupling implies that such a black hole solution cannot be trusted as it is beyond the regime of validity of the EFT of ghost-free higher-derivative theory.

For a class of stealth solutions with $X={\rm const.}$ and linearly time-dependent scalar field configurations, it was recently pointed out that one of the even-parity modes of liner perturbations has a vanishing sound speed, that the effective metric on which the mode propagates is singular and that as a result the mode is infinitely strongly coupled~\cite{deRham:2019gha}. In particular, a sufficient condition for the sound speed to vanish was derived.\footnote{For this analysis, one carefully performs a coordinate redefinition to diagonalize the time and spatial derivative terms and then checks the un/boundedness of the Hamiltonian~\cite{Babichev:2017lmw,Babichev:2018uiw}. Taking into account this point, stability analysis of static, spherically symmetric spacetime with a linearly time-dependent scalar field in DHOST theories can be formulated for the odd-parity perturbations \cite{Takahashi:2019oxz}. While the full stability analysis of even-parity perturbations has still been not clarified yet, 
\cite{deRham:2019gha,Charmousis:2019fre} 
succeeded in extracting the equation of motion of one of the even mode without the full analysis, and 
\cite{deRham:2019gha} obtained the sufficient condition for the strong coupling.}
Therefore, the results obtained in \cite{Minamitsuji:2018vuw,deRham:2019gha} for spacelike and timelike profiles of the scalar field respectively show that for these two types of stealth solutions the second even mode is infinitely strongly coupled, and hence they cannot be trusted. 
Of course, these results do not exclude the possibility of other stealth solutions without strong coupling.
Indeed, there exist other stealth solutions~\cite{Motohashi:2019sen,Takahashi:2019oxz} in DHOST theories that violate the sufficient condition for the vanishing sound speed, for which an independent study is required.

The two types of stealth solutions exhibit the infinite strong coupling not only in the bulk of the geometry but also in the asymptotic region, where the metric approaches either Minkowski or de Sitter. Therefore, let us consider a stealth Minkowski or de Sitter solution, hoping that it provides a hint for a possible solution of the problem. Here, for simplicity we consider a stealth Minkowski solution with a constant, timelike derivative of the scalar field, i.e.\ ghost condensation~\cite{ArkaniHamed:2003uy}. The EFT describing perturbations around the stealth Minkowski solution can be constructed systematically (see section 6 of \cite{ArkaniHamed:2003uy}). In the decoupling limit it is 
\begin{equation}
 S = \frac{1}{2}M^4 \int dt d^3\vec{x} \left[ \dot{\pi}^2 - \frac{\alpha}{M^2}(\vec{\nabla}^2\pi)^2  + \cdots \right]\,, \label{eqn:EFT-Minkowski-decoupling}
\end{equation}
where we have chosen the time coordinate $t$ so that the background scalar field has the form $\phi = M^2 t$, $\pi$ is the (rescaled) perturbation of the scalar field, $\alpha$ is a constant of order unity and dots represent nonlinear interactions of $\pi$. This shows an universal dispersion relation of the form $\omega^2 = \alpha k^4/M^2$ without the usual $k^2$ term. In order to estimate the energy dependence of the nonlinear interactions, let us first determine the scaling dimension of $\pi$ as the energy scale $E$ is scaled as $E\to sE$ (and thus $dt \to s^{-1}dt$), where $s$ is a constant. The dispersion relation implies that $d\vec{x} \to s^{-1/2}d\vec{x}$ under the scaling. By requiring that the quadratic part of the above action be invariant under the scaling, one concludes that $\pi$ should scale as $\pi \to s^{1/4}\pi$, meaning that the scaling dimension of $\pi$ is not $1$ but $1/4$. This then makes it possible to estimate the scaling dimensions of any nonlinear operators. For example, the leading nonlinear operator $M^4\int dt d\vec{x} (\vec{\nabla}\pi)^2\dot{\pi}$ scales as $s^{1/4}$ and thus is suppressed at low energy as $\sim (E/M)^{1/4}$. Similarly, one can check that any nonlinear operators are suppressed by $\sim (E/M)^{1/4}$ or higher order in $E/M$, meaning that the theory is weakly coupled all the way up to the scale $M$, as far as dimensionless parameters in the action are of order unity.

The structure of the EFT action (\ref{eqn:EFT-Minkowski-decoupling}) is determined solely by the symmetry breaking pattern and the derivative expansion, and thus can describe the low energy behavior of a wide class of underlining theories. However, if the underlining theory has a specific structure that forces the perturbations to follow second-order equations of motion (this is the case in DHOST theories) then $\alpha=0$ and the dispersion relation in the decoupling limit no longer depends on the spatial momentum as the $k^2$ term is forbidden by the symmetry. This explains the origin of the infinite strong coupling problem around stealth solutions in DHOST theories. This, at the same time, gives a hint for a solution of the problem: one can slightly detune the degeneracy condition to introduce higher spatial derivative terms suppressed by some high scale. A new term contributes to the quadratic action, bringing back the dispersion relation to the universal form $\omega^2\simeq \alpha k^4/M^2$ since there is no symmetry reason to forbid the $k^4$ term. The detuning also introduces higher time derivative terms and thus apparent ghost degrees of freedom. However, as far as the amount of detuning is small enough, those apparent extra modes have large masses above the cutoff and, as already stressed in the second paragraph of this section, should not be considered as physical. Therefore, a controlled detuning of the degeneracy condition is expected to solve the infinite strong coupling problem of the stealth solutions in DHOST theories. We name this mechanism as {\it scordatura}, an Italian word literally meaning ``detuning'', after a non-standard tuning of string musical instruments intended for making special chords possible and/or certain passages easier to play than the standard tuning.

The corrections to the metric due to the higher dimensional operators are expected to be unobservably small at astrophysical scales. This can be shown explicitly in ghost condensation, which admits approximately stealth black hole solutions. As shown in \cite{Mukohyama:2005rw}, the deviation of the metric from the corresponding GR solution is suppressed by the ratio $M^2/M_{\rm Pl}^2$ with $M \lesssim 100 \ {\rm GeV}$~\cite{ArkaniHamed:2005gu}, and thus is unobservably small at astrophysical scales.\footnote{The same conclusion holds for gauged ghost condensation~\cite{Cheng:2006us}.} On the other hand, as we have seen above at least in the asymptotic flat region, the perturbation around the solution is weakly coupled all the way up to the cutoff scale $M$. Hence, the approximately stealth black hole solution in ghost condensation is stealth for all practical purposes\footnote{Conceptually, on the other hand, the deviation from the corresponding GR solution is important for the recovery of the generalized second law~\cite{Mukohyama:2009rk,Mukohyama:2009um}.} and do not suffer from the above mentioned problem of the infinite strong coupling. We expect the same in scordatura DHOST theories. Namely, we expect that, as far as the amount of detuning of the degenerate condition is under control, stealth black hole solutions in scordatura DHOST theories are stealth for all practical purposes and do not suffer from the above mentioned problem of infinite strong coupling.

In this paper, we focus on the stealth solution with timelike derivative of the scalar field. (As we shall show later in this paper, the scordatura mechanism does not work if the derivative of the scalar field is spacelike.) We first point out that the strong coupling problem is universal and thus unavoidable, at least in the decoupling limit in the asymptotic region, where the background approaches stealth Minkowski or de Sitter solution. Away from the decoupling limit, the dispersion relation may receive corrections suppressed by negative powers of $M_{\rm Pl}^2$. However, because of the Planckian suppression, those corrections are small and the strong coupling scale of the perturbation remains too low for the EFT to be useful for interesting applications. We then show that the problem can be cured by scordatura, i.e.\ a controlled detuning of the degeneracy with introduction of additional higher-derivative terms. 
Therefore, stealth solutions in scordatura degenerate theories should be free from the problem of infinite strong coupling, provided that the derivative of the scalar field is timelike.

The rest of the paper is organized as follows. 
In \S\ref{sec:scs}, we estimate the energy scale of strong coupling in the asymptotic region of stealth solutions, where the geometry approaches either Minkowski or de Sitter, based on EFT of inflation in the decoupling limit. We also derive the dispersion relation in the decoupling limit, leaving the full analysis beyond the decoupling limit to Appendix~\ref{sec:lin}. 
In \S\ref{sec:ddh}, we focus on a stealth solution of a specific class of a scordatura DHOST theory.
We shall see that without the scordatura term, either strong coupling or gradient instability is inevitable for this class of theory. To be more precise, without the scordatura term we shall see that the strong coupling scale is 
much lower than $M$
and that the sound speed squared is negative and of order $\mathcal{O}(M^2/M_{\rm Pl}^2)$. Here, it is supposed that the action of the system in the decoupling limit is parameterized by the scale $M$ and dimensionless parameters of order unity.
Therefore, the strong coupling scale is rather low (suppressed by some powers of $M^2/M_{\rm Pl}^2 \ll 1$) and the system exhibits gradient instability in the rather narrow window below this low strong coupling scale, if the scordatura mechanism is not employed.
We then show that an introduction of the scordatura term cures the issue.  
In \S\ref{sec:conc}, we discuss our results and outlook.

\section{Simple estimates of strong coupling scales}
\label{sec:scs}

In this section, we consider general theory with timelike derivative of the scalar field in the Einstein frame 
that respects spatial diffeomorphism invariance in a Friedmann-Lema\^itre-Robertson-Walker (FLRW) background.
In particular, our treatment includes the case of the flat chart of de Sitter spacetime as a special case.
The discussion here is based on the EFT of single-field inflation~\cite{Creminelli:2006xe,Cheung:2007st}, 
which extends the EFT of ghost condensation, developed in section 6 of \cite{ArkaniHamed:2003uy} (see also \cite{ArkaniHamed:2003uz}), to a general FLRW inflationary background.

In general, for the EFT to cover a wide class of theories of modified gravity, one needs a further generalization of the EFT action to include additional terms~\cite{Gleyzes:2013ooa,Kase:2014cwa,Gleyzes:2014rba,Gleyzes:2015pma,Motohashi:2017gqb,Langlois:2017mxy,Motohashi:2020wxj}.
In particular, Horndeski theory generates a peculiar interaction between curvature and lapse, and general DHOST theories include time derivative of lapse.
However, we restrict our analysis to the canonical EFT action, since requiring the compatibility to gravitational wave observations, the additional EFT interactions should vanish~\cite{Creminelli:2018xsv}.
The remaining subclass of Horndeski or DHOST theories can be recast to the Einstein frame action through conformal and/or disformal transformations.
These transformations simply changes a de Sitter solution with $X={\rm cosnt.}$ to another de Sitter solution with $X={\rm cosnt.}$ as it merely causes constant rescaling of the lapse and scale factor.
The scalar field profile is also unchanged.

Another subtlety is about the chart of de Sitter spacetime.
We focus on asymptotically de Sitter stealth solutions with linearly time dependent scalar field, and consider the limit of spatial infinity.
The asymptotic form of the Schwarzschild--de Sitter or Kerr--de Sitter metric is given by the static chart of de Sitter spacetime, which can be transformed to the flat chart of de Sitter spacetime.
Under this transformation, the radial dependency of the scalar field can be removed
for a certain class of solutions, and hence we can regard it as the unitary gauge.
Specifically, for the Schwarzschild--de Sitter solution in the shift-symmetric quadratic DHOST theories, such a class is Case 1-$\Lambda$ identified in \cite{Motohashi:2019sen}.
We provide more detailed argument in Appendix~\ref{app:dS}.

Therefore, the de Sitter limit of the following EFT analysis in the unitary gauge can be regarded as the limit of spatial infinity of asymptotically de Sitter stealth solutions of Case 1-$\Lambda$ in Class Ia of DHOST theories, including Horndeski/GLPV subclass, with timelike scalar field.
Using this framework, below we focus on the decoupling limit action to estimate the strong coupling scale and 
show that at the de Sitter limit the dispersion relation of Nambu--Goldstone mode contains $k^2$ and $k^4$ term.
The linear perturbation analysis away from the decoupling limit in Minkowski limit is provided in Appendix~\ref{sec:lin}.

\subsection{EFT action}
\label{sec:eft}

Assuming the existence of the timelike scalar field, we take the unitary gauge so that the scalar field is given by $\bar \phi=t$.
The perturbation of the scalar field vanishes $\delta\phi=0$ by definition.
The residual gauge degree of freedom is purely spatial transformation, $\vec{x}\to \vec{x}'= \vec{x}'(t,\vec{x})$.
In the unitary gauge the EFT action which respects spatial diffeomorphism invariance and describes the perturbation of general theory in the Einstein frame around the FLRW spacetime is given by
\begin{equation} \label{EFTaction}
 S = \Mpl^2\int d^4x\sqrt{-g}
  \left[\frac{1}{2}R + c_1(t) + c_2(t)g^{00}
   + 
   \L^{(2)}(\tilde{\delta}g^{00}, \tilde{\delta}K_{\mu\nu}, 
   \tilde{\delta}R_{\mu\nu\rho\sigma}; 
   t,  g_{\mu\nu}, g^{\mu\nu}, \nabla_{\mu})
  \right]\,, 
\end{equation}
where 
\begin{equation}
 \L^{(2)} =  
  \lambda_1(t)(\tilde{\delta}g^{00})^2
  + \lambda_2(t)(\tilde{\delta}g^{00})^3
   + \lambda_3(t)\tilde{\delta}g^{00}\tilde{\delta}K^{\mu}_{\mu}
   + \lambda_4(t)(\tilde{\delta}K^{\mu}_{\mu})^2
   + \lambda_5(t)\tilde{\delta}
   K^{\mu}_{\nu}\tilde{\delta}K_{\mu}^{\nu}
   + \cdots\,,
\end{equation}
and
\begin{eqnarray}
 \tilde{\delta}g^{00}& \equiv &
  g^{00}+1\,,  \quad
 \tilde{\delta}K_{\mu\nu} \equiv
  K_{\mu\nu} - H \gamma_{\mu\nu}\,, \nonumber\\
 \tilde{\delta}R_{\mu\nu\rho\sigma}
 & \equiv &
  R_{\mu\nu\rho\sigma}
  - 2(H^2+{\cal K}/a^2) \gamma_{\mu[\rho}\gamma_{\sigma]\nu}
  + (\dot{H}+H^2) (\gamma_{\mu\rho}\delta^0_{\nu}\delta^0_{\sigma} + (3 \mbox{perm.}))\,. 
\end{eqnarray}
Here, $n_{\mu} \equiv \delta_{\mu}^0/\sqrt{-g^{00}}$ is the unit normal vector, $\gamma_{\mu\nu} \equiv g_{\mu\nu} + n_{\mu}n_{\nu}$ is the induced metric, and $\mathcal{K}/a^2$ is the spatial curvature of the FLRW background geometry with the scale factor $a(t)$. The background lapse function is set to be unity so that the Hubble expansion rate for the background is $H=\dot{a}/a$ with a dot denoting a derivative with respect to $t$. 
The mass dimension of the $\lambda_i$ functions are $\lambda_1, \lambda_2$: 2, $\lambda_3$: 1, and $\lambda_4, \lambda_5$: 0.

The terms linear in the perturbation give the background equations of motion as
\be
 3H^2+\frac{3{\cal K}}{a^2} = -c_1-c_2\,,  \quad 
  \dot{H}-\frac{\cal K}{a^2} = c_2\,.
\ee
Solving the background equations of motion with respect to $c_1$ and $c_2$ and plugging them back into the action~\eqref{EFTaction}, we obtain
\begin{equation}
 S = \Mpl^2\int d^4x\sqrt{-g}
  \left[\frac{1}{2}R -\left(3H^2+\dot{H}+\frac{2{\cal K}}{a^2}\right)
   + \left(\dot{H}-\frac{\cal K}{a^2}\right)g^{00} + \L^{(2)} \right]\,.
  \label{eqn:unitary-gauge-action}
\end{equation}

\subsection{Decoupling limit action}
\label{sec:dec}

Leaving the full analysis taking metric perturbations into account in Appendix~\ref{sec:lin}, 
for the rest of \S\ref{sec:scs} we focus on the decoupling limit action
neglecting metric perturbations.
This treatment dramatically simplifies the analysis, and is justified on sufficiently small scales~\cite{Cheung:2007st}.
In the unitary gauge action~\eqref{eqn:unitary-gauge-action}, perturbation of scalar field does not appear explicitly but is encoded in the metric perturbations.
By following St\"uckelberg trick, we can obtain the quadratic action for the Nambu--Goldstone mode $\pi$ at the decoupling limit.
By acting the broken time diffeomorphism $t\to t'=t-\pi(t', \vec{x})$ 
on the unitary gauge action (\ref{eqn:unitary-gauge-action}) and then rewriting $t'$ as $t$, one obtains the covariant EFT action. 
In practice, one only needs to make the following replacements in (\ref{eqn:unitary-gauge-action})
\begin{eqnarray}
 \delta^0_{\mu} & \to &
  (1+\dot{\pi})\delta^0_{\mu}
  + \delta^i_{\mu}\partial_i\pi\,,  \nonumber\\
 H(t) & \to & H(t+\pi)\,,  \quad
 \dot{H}(t) \to \dot{H}(t+\pi)\,,\quad 
 \lambda_i(t) \to \lambda_i(t+\pi)\,,   \quad
  a(t) \to a(t+\pi)\,,  \nonumber\\
 g_{\mu\nu} & \to & g_{\mu\nu}\,,  \quad
 g^{\mu\nu} \to g^{\mu\nu}\,,  \quad
  \nabla_{\mu} \to \nabla_{\mu}\,,  \quad
   R_{\mu\nu\rho\sigma} \to  R_{\mu\nu\rho\sigma}\,. 
   \label{eqn:broken-time-diffeo}
\end{eqnarray}
In the decoupling limit one neglects the metric perturbations to obtain the action for the Nambu--Goldstone mode $\pi$ as
\be \label{Spi}
S_{\pi} = \Mpl^2\int dt d^3\vec{x}\, a^3
\left[  (-\dot H+4\lambda_1) \dot\pi^2 + \dot H \f{(\pa_i\pi)^2}{a^2} 
+ 4(\lambda_1-2\lambda_2)\dot\pi^3 - 4\lambda_1 \dot\pi \f{(\pa_i\pi)^2}{a^2} 
+ \O(\pi^4, \tilde{\epsilon}^2)
+ \L^{(2)}_{\tilde{\delta}K, \tilde{\delta}R} \right] ,
\ee
where we have set $\mathcal{K}=0$ for simplicity, 
we have assumed the adiabatic evolutions of $H$ and $\lambda_i$ as
\begin{equation}
 \left|\frac{(\partial_t)^n H}{H^{n+1}}\right| = \O(\tilde{\epsilon}^n)\,, \quad
  \left|\frac{(\partial_t)^n\lambda_i}{H^n\lambda_i}\right|
  = \O(\tilde{\epsilon}^n)\,,  \quad
  |\tilde{\epsilon}|\ll 1\,,  \quad n=1, 2, \cdots\,,
  \label{eqn:slow-roll-cond}
\end{equation} 
and $\L^{(2)}_{\tilde{\delta}K, \tilde{\delta}R}$ represents those terms in $\L^{(2)}$ that involve $\tilde{\delta}K_{\mu\nu}$ and/or $\tilde{\delta}R_{\mu\nu\rho\sigma}$ such as $\lambda_3, \lambda_4, \lambda_5$ terms and that depend on higher derivatives of $\pi$.

Below we estimate the energy scale $E_{\rm cubic}$, at which the cubic terms become comparable to the quadratic kinetic terms, for each case 
where $|-\dot H/(4\lambda_1)|$ is not too small, or $|-\dot H/(4\lambda_1)| \ll 1$.
$E_{\rm cubic}$ gives an upper bound of the strong coupling scale.

\subsubsection{Case with not-too-small $c_{\rm s}^2$}
\label{sec:EFTinflation}

For the case where $|-\dot H/(4\lambda_1)|$ is not too small, the decoupling limit action~\eqref{Spi} can be rewritten as
\begin{eqnarray}
 S_{\pi} & = & \Mpl^2\int dt d^3\vec{x}\,  a^3
  \left[
   - \frac{\dot{H}}{c_{\rm s}^2}
   \left(\dot{\pi}^2-c_{\rm s}^2\frac{(\partial_i\pi)^2}{a^2}\right)
   -\dot{H}\left(\frac{1}{c_{\rm s}^2}-1\right)
   \left(\frac{c_3}{c_{\rm s}^2}\dot{\pi}^3
    -\dot{\pi}\frac{(\partial_i\pi)^2}{a^2}\right)
      + \O(\pi^4, \tilde{\epsilon}^2)
      + \L^{(2)}_{\tilde{\delta}K, \tilde{\delta}R} \right]\,, 
  \label{eqn:pi-action}
\end{eqnarray}
where 
we have introduced $c_{\rm s}^2$ and $c_3$ by
\begin{equation}
 \frac{1}{c_{\rm s}^2} = 1 + \frac{4\lambda_1}{-\dot{H}}\,,  \quad
 c_3 = c_{\rm s}^2 - \frac{8c_{\rm s}^2\lambda_2}{-\dot{H}}
  \left(\frac{1}{c_{\rm s}^2}-1\right)^{-1}\,.
  \label{eqn:def-cs2-c3}
\end{equation}
Therefore, if $|-\dot H/(4\lambda_1)|$ is not too small, so is $c_{\rm s}^2$.
If $c_{\rm s}^2$ is not too small then one can safely ignore the effects of $\L^{(2)}_{\tilde{\delta}K, \tilde{\delta}R}$.

To estimate the strong coupling scale, for simplicity we further assume that $c_{\rm s} \simeq {\rm const.}$ in the time scale of order $1/E$, where $E$ is the energy scale of interest, and rescale the spatial coordinates as
\begin{equation}
\vec{x} = c_{\rm s} \vec{\tilde{x}}\,.
\end{equation}
We then obtain
\begin{eqnarray}
 S_{\pi} & = & \int dt d^3\vec{\tilde{x}}\,  a^3 (c_{\rm s}\epsilon \Mpl^2 H^2)
  \left[
   \dot{\pi}^2-\frac{(\tilde{\partial_i}\pi)^2}{a^2}
   + \left(\frac{1}{c_{\rm s}^2}-1\right)\dot{\pi}
   \left(c_3 \dot{\pi}^2
    -\frac{(\tilde{\partial_i}\pi)^2}{a^2}\right) + \cdots \right]\,,  \label{eqn:pi-action-rescaled}
\end{eqnarray}
where $\epsilon \equiv -\dot{H}/H^2$ is the slow-roll parameter. From now on we assume that $0<c_{\rm s}^2<1$. Avoidance of strong coupling requires that the first two terms dominate over the nonlinear terms. With this condition one can estimate the amplitude of quantum fluctuations for a given energy scale $E$ as
\begin{equation}
 \dot{\pi}^2 \sim \frac{(\tilde{\partial_i}\pi)^2}{a^2} \sim \frac{E^4}{c_{\rm s}\epsilon \Mpl^2 H^2}\,.
  \label{eqn:amplitude}
\end{equation}
We now would like to estimate the energy scale $E_{\rm cubic}$ at which the cubic terms become comparable with the quadratic terms, i.e.
\begin{equation}
 \left. \left(\frac{1}{c_{\rm s}^2}-1\right) |\dot{\pi}|\, \right|_{E = E_{\rm cubic}} \sim \frac{1}{\max [|c_3|,1]}\,.
  \label{eqn:def-cubicstrongcoupling}
\end{equation}
Combining (\ref{eqn:amplitude}) and (\ref{eqn:def-cubicstrongcoupling}), one can estimate $E_{\rm cubic}$ as
\begin{equation} \label{eqn:Ecubic}
 E_{\rm cubic} \lesssim \frac{(c_{\rm s}^5\epsilon \Mpl^2 H^2)^{1/4}}{\sqrt{1-c_{\rm s}^2}}\,,
\end{equation}
where we have assumed that the first two terms in (\ref{eqn:pi-action-rescaled}) remain to be the dominant quadratic terms all the way up to $E\sim E_{\rm cubic}$. Obviously, the strong coupling scale is lower than or equal to $E_{\rm cubic}$ since higher order terms may or may not become comparable with the quadratic terms below $E_{\rm cubic}$. Therefore, if the first two terms in (\ref{eqn:pi-action-rescaled}) remain to be the dominant quadratic terms then the system would be infinitely strongly coupled in the limit $c_{\rm s}^5\epsilon/(1-c_{\rm s}^2)^2 \to 0$. 
However, in this limit, $c_{\rm s}\ll 1$ or $|-\dot H/(4\lambda_1)|\ll 1$, terms in $\L^{(2)}_{\tilde{\delta}K, \tilde{\delta}R}$ are not negligible and need to be taken into account, which we shall address below.

\subsubsection{de Sitter limit ($c_{\rm s}^2\ll 1$)}
\label{ssec:dslim}

Let us focus on the case where $|-\dot H/(4\lambda_1)| \ll 1$ and estimate the strong coupling scale.
In this case the background expansion is close to de Sitter $H\simeq {\rm const}$, and the sound speed is small $c_{\rm s}^2 \ll 1$.
This limit is not necessarily a fine-tuning since the de Sitter limit 
can be naturally realized as an attractor of a system~\cite{ArkaniHamed:2003uy}.

As stated just after \eqref{eqn:def-cs2-c3}, in the previous subsection we have ignored $\L^{(2)}_{\tilde{\delta}K, \tilde{\delta}R}$, assuming that $c_{\rm s}^2$ is not too small. 
In the case $c_{\rm s}^2\ll 1$ or de Sitter limit, 
extra terms hidden in $\L^{(2)}_{\tilde{\delta}K, \tilde{\delta}R}$ of (\ref{eqn:pi-action}) cannot be ignored.
With $c_{\rm s}^2\ll 1$ the EFT action~\eqref{Spi} for $\pi$ at the decoupling limit reads
\begin{align}
 S_{\pi} = \Mpl^2\int dt d^3\vec{x}\,  a^3
  \Bigg[
   &4\lambda_1 \left(\dot{\pi}^2
	       -c_{\rm s}^2\frac{(\partial_i\pi)^2}{a^2}
	       - \dot{\pi}\frac{(\partial_i\pi)^2}{a^2}
       \right)
 + 4(\lambda_1-2\lambda_2)\dot{\pi}^3 \notag\\
  &+\lambda_3 \left( H-\frac{\partial_j^2\pi}{a^2}\right)
  \frac{(\partial_i\pi)^2}{a^2}
  + (\lambda_4+\lambda_5)\frac{(\partial_i^2\pi)^2}{a^4}
  + \cdots
  \Bigg]\,,
  \label{eqn:pi-action-dS-cs}
\end{align} 
where we 
performed integration by parts
and neglected subdominant terms under $H/M\ll 1$, $\omega/M\ll 1$, $k/(Ma) \ll 1$, otherwise the EFT would be useless.

Let us employ the notation 
\be \label{deflam}
\lambda_1 = \f{M^4}{8 \Mpl^2}, \quad 
\lambda_3 = \f{M^3 \beta}{2 \Mpl^2}, \quad 
\lambda_4 = -\f{M^2 (\alpha + \gamma)}{2 \Mpl^2}, \quad 
\lambda_5 = \f{M^2 \gamma}{2 \Mpl^2}.  \ee
The action~ \eqref{eqn:pi-action-dS-cs} can then be rewritten as
\begin{align}
  S_{\pi} = \f{M^4}{2} \int dt d^3\vec{x}\,  a^3
   \Bigg[
    \dot{\pi}^2
          -c_{\rm s}^2\frac{(\partial_i\pi)^2}{a^2}
          - \dot{\pi}\frac{(\partial_i\pi)^2}{a^2}
   -\f{\alpha}{M^2} \frac{(\partial_i^2\pi)^2}{a^4}
   +\f{\beta}{M} \left( H-\frac{\partial_j^2\pi}{a^2}\right)
   \frac{(\partial_i\pi)^2}{a^2}
   + \cdots
   \Bigg]\,,
\end{align}

Let us estimate the energy scale $E_{\rm cubic}$ at which the cubic term $\dot{\pi}(\partial_i\pi)^2a^{-2}$ becomes comparable to the canonical kinetic term $\dot{\pi}^2$ which we assume is order unity.
First, requiring order unity kinetic term part of the action yields $E^{-1}p^{-3}M^4(E\pi)^2 \sim 1$, namely,
the amplitude of quantum fluctuations for given energy scale $E$ and physical momentum scale $p$ is estimated as
\be \label{pieq} \pi \sim \f{E^{3/2}}{p^{1/2}M^2}. \ee
Second, requiring cubic term comparable to the quadratic term yields 
\be \label{Eeq} \left. \f{E \pi p^2}{E^2}\right|_{E=E_{\rm cubic}} \sim 1 . \ee
Combining \eqref{pieq} and \eqref{Eeq}, we obtain the following equation that should be satisfied at $E=E_{\rm cubic}$
\be \label{Ecubic} \left. \mk{\f{p}{E}}^{7/4} \f{E}{M} \right|_{E=E_{\rm cubic}} \sim 1 . \ee
To explicitly write down $E_{\rm cubic}$ we need a relation between $E$ and $p$.
Below the strong coupling scale the quadratic terms in \eqref{eqn:pi-action-dS-cs} should dominate over the nonlinear terms
and the physical momentum $p$ is related to $E$ through the dispersion relation obtained by the quadratic terms.

We assume de Sitter limit $c_{\rm s}^2\ll 1$ and the EFT assumption 
$H/M\ll 1$, $\omega/M\ll 1$, $k/(Ma) \ll 1$, and $|\alpha|, |\beta| = \mathcal{O}(1)$.
The dispersion relation for general case is then given by
\be \label{disp-gen} \f{\omega^2}{M^2} = \mk{c_{\rm s}^2-\beta \f{H}{M} } \f{k^2}{M^2a^2} + \alpha \f{k^4}{M^4a^4}. \ee
Thus, the dispersion relation varies depending on which term on the right hand side is dominant.
As we shall see below, there is a crucial difference of the estimation of $E_{\rm cubic}$ between the case where $\alpha$ term is dominant and other cases.

First, let us consider the case where $c_{\rm s}^2$ term is dominant,
\be {\rm max} \kk{ |\alpha| \f{k^2}{M^2a^2}, |\beta| \f{H}{M} } \ll c_{\rm s}^2 \ll 1, \ee 
which requires non-exact de Sitter spacetime and includes the case $\alpha=0$.
In this case \eqref{disp-gen} reads 
\be \label{disp-1} \f{\omega^2}{M^2} = c_{\rm s}^2 \f{k^2}{M^2a^2}. \ee
It means $E\simeq c_{\rm s}p$, with which from \eqref{Ecubic} we can estimate $E_{\rm cubic}$ as
\be \label{Ecubic-1} E_{\rm cubic} \simeq c_{\rm s}^{7/4} M \ll M . \ee
Therefore, the strong coupling scale is much lower than $M$.
In this case we have neglected terms originating from $\L^{(2)}_{\tilde{\delta}K, \tilde{\delta}R}$, and hence \eqref{Ecubic-1} is consistent with \eqref{eqn:Ecubic} in the limit $c_{\rm s}^2\ll 1$.

Next, let us consider the case where $\beta $ term is dominant,
\be {\rm max} \kk{ c_{\rm s}^2 , |\alpha| \f{k^2}{M^2a^2} } \ll  |\beta| \f{H}{M}\ll 1, \ee 
which includes the case $\alpha=0$ and/or exact de Sitter.
In this case \eqref{disp-gen} reads 
\be \label{disp-2} \f{\omega^2}{M^2} = -\beta \f{H}{M} \f{k^2}{M^2a^2}, \ee
meaning $E\simeq (|\beta| H/M)^{1/2}p$, with which from \eqref{Ecubic} we obtain 
\be \label{Ecubic-2} E_{\rm cubic} \simeq \mk{|\beta| \f{H}{M}}^{7/8} M \ll M . \ee
Again, the strong coupling scale is much lower than $M$.

Finally, let us consider the case where $\alpha$ term is dominant,
\be {\rm max} \kk{ c_{\rm s}^2 , |\beta| \f{H}{M} } \ll |\alpha| \f{k^2}{M^2a^2} \ll 1, \ee 
which is possible if $\alpha \ne 0$, and
includes the case of exact de Sitter.
In this case \eqref{disp-gen} reads 
\be \label{disp-3} \f{\omega^2}{M^2} = \alpha \f{k^4}{M^4a^4}, \ee
meaning $E\simeq |\alpha|^{1/2}p^2/M$, with which from \eqref{Ecubic} we obtain 
\be \label{Ecubic-3} E_{\rm cubic} \simeq |\alpha|^{7/2} M. \ee

Unlike the first two cases \eqref{Ecubic-1} and \eqref{Ecubic-2}, in the last case case the strong coupling scale~\eqref{Ecubic-3} can be as high as $M$ provided that $\alpha$ is of order unity.
This is consistent with the estimation of the scaling dimensions of nonlinear operators which we argued below \eqref{eqn:EFT-Minkowski-decoupling}.
The EFT action for $\pi$ in this regime is
\begin{equation}
 S_{\pi} = \int dt d^3\vec{x}\,  a^3\,  \frac{M^4}{2}
  \left[\dot{\pi}^2 - \frac{\alpha}{M^2}\frac{(\partial_i^2\pi)^2}{a^4}
	       - \dot{\pi}\frac{(\partial_i\pi)^2}{a^2} + \cdots \right]\,,
  \label{eqn:pi-action-gc}
\end{equation}
where the first two terms are dominant.
This is precisely an analogous equation to \eqref{eqn:EFT-Minkowski-decoupling}.
The stability of $\pi$ requires that $M^4>0$ and $\alpha>0$.

There are several remarks on the above analysis.
As mentioned above, a caveat for the decoupling limit analysis is that it is valid only on sufficiently small scales.
This is because in general quadratic terms with $h_{\mu\nu}$ and $\pi$ have fewer derivatives than the kinetic term of $\pi$, and hence they can be neglected only above some energy scale.  
Therefore, it is inevitable for the decoupling limit analysis to have an ambiguity for terms in lower order in $k$.

Another point about the $\O(k^0)$ term is that for expanding universe the contribution from superhorizon mode at $\O(k^0)$ depends on a choice of coordinate or gauge.
Therefore, we focus stability of subhorizon modes, i.e.\ terms in higher order of $k$.

As a complementary analysis taking into account these subtleties, in Appendix~\ref{sec:lin} we provide an analysis beyond the decoupling limit by including metric perturbations.
It is found that
the corrections to the dispersion relation
is of order $\mathcal{O}({M^2/\Mpl^2)}$ (see \eqref{disp-eftlin}).
Also, we take the Minkowski limit to avoid the ambiguity of a choice of gauge, so we can discuss stability including terms in lower order in $k$.

The above analysis suggests the possibility to fix strong coupling 
for $k^2$ term by taking into account $k^4$ term. 
However, so long as ones considers a Lorentz invariant theory satisfying degeneracy condition, such as DHOST theories, one would not have $k^4$ term in the dispersion relation since such a theory is essentially governed by (temporal and spatial) second-order differential equations.
The $k^4$ term appears if we consider detuning of such a theory, by introducing either of Lorentz violating term or higher-derivative term, which violates the degeneracy condition.
We shall investigate this realization in a specific class of a scordatura DHOST theory in \S\ref{sec:ddh}.

\subsection{Timelike vs spacelike}
\label{ssec:tmsp}

Finally, before closing this section, we remark that the above logic does not hold if $\pa_\mu \phi$ is spacelike.  
For simplicity, let us consider the Minkowski limit of the analaysis in \S\ref{ssec:dslim}.

First, to reiterate, with timelike and constant $\partial_{\mu}\phi$ in the Minkowski background, we have shown in \eqref{eqn:pi-action-gc} that the leading-order quadratic action for the Nambu--Goldstone boson is of the form
\begin{equation}
 S_\pi^{(2)} = \frac{M^4}{2}\int dt d^3\vec{x} \left[ \epsilon (\partial_t\pi)^2  - \frac{\alpha}{M^2}(\partial_x^2\pi+\partial_y^2\pi+\partial_z^2\pi)^2 \right]\,,
\end{equation}
where $\epsilon=\pm 1$, $\alpha$ is a dimensionless constant, $M$ is an energy scale and we have chosen the Lorentz frame so that $\partial_{\mu}\phi\propto\delta_{\mu}^t$ for the background. If one fine-tunes the parameters of the theory so that the equation of motion is second-order differential equation then $\alpha=0$ and the theory is strongly coupled at all scales. On the other hand, for $\epsilon=1$ and a positive $\alpha$ of $\mathcal{O}(1)$, from \eqref{Ecubic-3} the strong coupling scale is of order $M$ and thus the theory is under a good theoretical control. In this way the strong coupling problem of the fine-tuned ($\alpha=0$) theory can be easily and consistently cured by simply relaxing the fine-tuning, if the condensate of $\partial_{\mu}\phi$ is timelike.

In contrast, if we instead consider a Minkowski background with spacelike and constant $\partial_{\mu}\phi$ then the leading-order quadratic action for the Nambu--Goldstone boson would be of the form
\begin{equation}
 S_\pi^{(2)} = \frac{M^4}{2}\int dt d^3\vec{x} \left[ \epsilon (\partial_z\pi)^2  - \frac{\alpha}{M^2}(-\partial_t^2\pi+\partial_x^2\pi+\partial_y^2\pi)^2 \right]\,,
\end{equation}
where 
we have chosen the Lorentz frame so that $\partial_{\mu}\phi\propto\delta_{\mu}^z$ for the background. Again, the fine-tuned theory with $\alpha=0$ is strongly coupled at all scales. On the other hand, for non-vanishing $\alpha$ of $\mathcal{O}(1)$  there always is a ghost without mass gap. Therefore, if the condensate of $\partial_{\mu}\phi$ is spacelike then one cannot cure the strong coupling problem of the $\alpha=0$ theory by relaxing the fine-tuning.

\section{Stealth solution in scordatura DHOST theories}
\label{sec:ddh}

In \S\ref{sec:scs} we employed EFT approach to explore general quadratic action in the Einstein frame with timelike scalar field, and clarified that to avoid the low strong coupling scale it is crucial to take into account $k^4$ term in the dispersion relation.
For covariant degenerate theory, such a term only shows up by considering detuning of covariance or degeneracy. 
In this section, as a concrete model we consider detuning or scordatura of a class of DHOST theory.
Considering the dispersion relation in the Minkowski limit, we shall see that the scordatura term $- \f{\alpha}{2M^2} (\Box\phi)^2$ precisely plays the role of resolving the strong coupling clarified in \S\ref{sec:scs}.

\subsection{Background}

Let us consider a scordatura DHOST theory 
\be \label{aciton} S = \int d^4x \sqrt{-g} \kk{ F_0 + F_1 \Box\phi + F_2 R + A_4 \phi^\mu \phi_{\mu\nu} \phi^{\nu\lambda} \phi_\lambda 
+ A_2 (\Box\phi)^2 }, \ee
where 
\be A_4=\f{6F_{2X}^2}{F_2} , \quad A_2= - \f{\alpha}{2M^2},
\ee
$F_0,F_1,F_2$ are functions of $X\equiv g^{\mu\nu}\phi_\mu\phi_\nu$, $\alpha$ is a dimensionless constant, and $M$ is a mass scale.
The last term with $\alpha$ is the scordatura term, or detuning term.
For $\alpha=0$, this model satisfies the degeneracy condition, and hence there is no Ostrogradsky ghost.
It also satisfies the conditions for $c_t=c$~\cite{Langlois:2017dyl} and no graviton decay~\cite{Creminelli:2018xsv}.
For $\alpha\ne 0$, the degeneracy condition is violated, so the model~\eqref{aciton} possesses an Ostrogradsky ghost, which shows up at the energy scale $M$.
So long as we consider the theory as an EFT up to the energy scale $M$, 
this apparent ghost degree of freedom is not a problem, since the properties and even the existence of the apparent ghost are sensitive to UV completion. 
Rather, since the degeneracy condition is not related to a symmetry in general, it is expected to be spoiled by quantum corrections, leading to apparent ghost degrees of freedom above a cutoff scale. 
Therefore, the scordatura DHOST model~\eqref{aciton} is a simple model of viable EFT with cutoff energy scale $M$.

This model allows stealth Schwartzschild--de Sitter solution of Cases 1-$\Lambda$ and 2-$\Lambda$~\cite{Motohashi:2019sen}.
Actually, for this model, Cases 1-$\Lambda$ and 2-$\Lambda$ conditions for the coupling functions are the same, and the remaining difference is only about the scalar field profile $\phi=qt+\psi(r)$: $X=-q^2$ for Case 1-$\Lambda$ and $X\ne -q^2$ for Case 2-$\Lambda$.

Let us introduce the following normalized quantities.  First, we denote 
$q = M^2$
and normalize coupling functions as
$F_{0} = f_0 M^4$,  
$F_{1} = M f_1$, 
$F_{2} = f_2 \Mpl^2$.
Derivatives are denoted as 
$F_{0,X} = f_{0x}$, $F_{0,XX} = f_{0xx}/M^4$, and so on.
We also normalize Hubble parameter as $H = h_0  M^2/\Mpl$ and assume that 
$\mu \equiv M/\Mpl \ll 1$.

Let us focus on the spatial infinity limit of the Case 1-$\Lambda$ stealth Schwarzschild--de Sitter solution with the scalar field profile $\phi=qt+\psi(r)$ with $X=-q^2$.  
At the spatial infinity, the Schwarzschild--de Sitter metric can be approximated by the static chart of de Sitter spacetime.
For the Case 1-$\Lambda$ solution, we can transform it to the one in the flat chart with $\phi=qt$ (see Appendix~\ref{app:dS}).
Therefore, as the background spacetime, we consider the flat chart of de Sitter spacetime~\eqref{flatdS} and work on the unitary gauge, i.e.\ $\bar \phi=qt$.
The background equations are then given by 
\begin{align} \label{bgeq}
f_0 &= -\f{3}{2} h_0^2 (4 f_2 + 3 \alpha \mu^2), \notag\\
f_{0x} &= 3 h_0 (-4 h_0 f_{2x} + \mu f_{1x}),
\end{align}
where the functions $f_i$ and their derivatives are evaluated at $X=-q^2$.
These equations coincide with those obtained for Schwarzschild--de Sitter solution in \cite{Motohashi:2019sen} when taking the de Sitter limit.

\subsection{Perturbations}
\label{sec:pert}

Let us consider perturbations around the flat chart of de Sitter spacetime~\eqref{flatdS}.
Since $F_2R$ is the only term involving the curvature, the no-ghost condition for tensor perturbations is simply given by
\be \label{noghost-ten} f_2 > 0 . \ee
The vector perturbation vanishes as in the standard case.
Below we focus on the scalar perturbation.
Working on the unitary gauge, the perturbation of the scalar field vanishes $\delta\phi=0$ by definition.
In general, scalar-perturbed flat FLRW metric is given by
\be \label{FLRWpert} ds^2 = - N^2(1+2\Phi)dt^2 + 2aN \pa_i Bdtdx^i + a^2\kk{(1+2\Psi)\delta_{ij} + \mk{ \pa_i\pa_j - \f{\triangle}{3} } E } dx^idx^j, \ee
where $N,a$ are the lapse function and the scale factor respectively, and
$\Phi, B, \Psi, E$ are perturbation variables.
Using the remaining gauge degree of freedom, we fix $E=0$.
Therefore, our gauge fixing condition is $\delta\phi=E=0$, which can be safely imposed at the action level since the metric and the scalar field profile share the same coordinate dependency and we do not lose any independent equation of motion~\cite{Motohashi:2016prk}.

We can integrate out nondynamical variables and reduce the quadratic Lagrangian $\L^{(2)}$ in the Fourier space as follows.
First, by using integration by parts, we remove terms such as $\ddot\Psi\Psi$, $\dot\Psi\Psi$, $\dot BB$, which are contained in $\L^{(2)}$.
Further integration by parts allows us to remove time derivative from $B$, while $\dot\Phi$ and $\dot\Psi$ remain in the Lagrangian.
At this stage, we arrive at the form 
\be \L^{(2)}=\L^{(2)}(\dot\Psi, \Psi, \dot\Phi, \Phi, B) . \ee
To further reduce the system, we introduce an auxiliary variable $Q$ and a Lagrange multiplier $\lambda$ to replace $\dot\Phi$ by $Q$:
\begin{align} \label{tL} 
\tilde \L^{(2)} &= \L^{(2)}(\dot\Psi, \Psi, Q, \Phi, B) + \lambda (Q-\dot\Phi)  \notag\\
&= \L^{(2)}(\dot\Psi, \Psi, Q, \Phi, B) + \lambda Q + \dot \lambda \Phi , 
\end{align}
where for the last line we used integration by parts. 
From \eqref{tL}, $\Phi, B, Q$ are clearly nondynamical variables.  
We can derive their constraint equations and solve them to express $\Phi, B, Q$ in terms of $\dot\Psi,\Psi,\dot\lambda,\lambda$. 
Substituting them back into $\tilde \L^{(2)}$, we obtain the quadratic Lagrangian depending on $\dot\Psi,\Psi,\dot\lambda,\lambda$. 
If $\alpha\ne 0$, we are left with two degrees of freedom, whereas if $\alpha=0$, we can further integrate out one nondynamical degree of freedom and arrive at the final Lagrangian with only one dynamical degree of freedom.

\subsubsection{Case $\alpha= 0$}
\label{sssec:1}

First, let us consider the standard DHOST case by setting $\alpha= 0$.
Since the full expressions are lengthy, from now on we shall demonstrate the reduction of $\tilde \L^{(2)}(\dot\Psi,\Psi,\dot\lambda,\lambda)$ at the Minkowski limit, which actually suffices our purpose.
For $\alpha=0$, we can complete the square the kinetic terms 
\be 
\tilde \L^{(2)} = -\f{f_2 }{ 2\Mpl^2 [ 4 f_{2x} (2 f_2 + 3 f_{2x}) k^2 + \Mpl^2 \mu^4 (4 f_{0xx} f_2 - 3 f_{1x}^2 \mu^2) ] } [\dot \lambda + 3 \Mpl^3 \mu^3 (f_1 - 2 f_{1x}) \dot\Psi]^2 + \cdots ,
\ee
where dots indicate terms at most linear order in $\dot \Psi$ and $\dot\lambda$.
We then define a new variable $\chi$ by 
\be \chi = \lambda + 3 \Mpl^3 \mu^3 (f_1 - 2 f_{1x}) \Psi , \ee
and erase $\lambda$ by substitution.
We can then integrate out $\Psi$, and further integration by parts leads us to the final Lagrangian 
\be \tilde \L^{(2)} = A [(4 f_{0xx} f_2 + 3 f_{1x}^2 \mu^2) \dot \chi^2 + f_{1x}^2 k^2 \mu^2 \chi^2] , \ee
where the overall factor is given by
\be A = \f{f_2 \mu^4}{32 f_{2x}^2 (f_2 + f_{2x})^2 k^4 - 8 f_{2x} k^2 \Mpl^2 \mu^4 [4 f_{0xx} f_2 f_{2x} + 3 f_{1x}^2 (2 f_2 + 3 f_{2x}) \mu^2] + 18 f_{1x}^4 \Mpl^4 \mu^{12}} . \ee
We see that at the leading order of $\mu\equiv M/\Mpl$, $A$ is proportional to $f_2$, which is positive from the no-ghost condition~\eqref{noghost-ten} for tensor perturbations.
Therefore, at the leading order of $\mu$,
the no-ghost condition for scalar perturbation is given by
\be \label{noghost-sca} f_{0xx} > 0 ,  \ee
and the sound speed is given by
\be c_{\rm s}^2 = - \f{f_{1x}^2 \mu^2}{4 f_{0xx} f_2} .  \ee
Note that the denominator is positive under the no-ghost conditions for tensor~\eqref{noghost-ten} and scalar~\eqref{noghost-sca}.
For $f_{1x}=0$, the sound speed vanishes and the system is strongly coupled as is studied in \S\ref{sec:scs}.
On the other hand, for $f_{1x}\ne 0$, the system suffers from gradient instability.
Therefore, either strong coupling or gradient instability is inevitable for the model with $\alpha=0$.
The negative sound speed 
squared is of order $\mathcal{O}(M^2/M_{\rm Pl}^2)$, which is consistent with the general dispersion relation \eqref{disp-eftlin} away from the decoupling limit.

\subsubsection{Case $\alpha\ne 0$}
\label{sssec:2}

Next, we shall show that the above issue can be cured if we take into account the scordatura term with nonzero $\alpha$.
For $\alpha \ne 0$, the system possesses two degrees of freedom.
One of them is the apparent Ostrogradsky ghost originating from the detuning of the degeneracy condition, which exists above the EFT cutoff energy scale.
The quadratic Lagrangian $\tilde \L^{(2)}(\dot\Psi,\Psi,\dot\lambda,\lambda)$ at the Minkowski limit is given by 
\be \tilde \L^{(2)} = \f{1}{2} \dot Y^T K \dot Y +  \dot Y^T M Y - \f{1}{2} Y^T W Y , \ee
where $Y^T = (\Psi,\lambda)$, $K$ and $W$ are $2\times 2$ symmetric matrices, and $M$ is $2\times 2$ antisymmetric matrix.
From the equations of motion, the dispersion relation is given by
\be \det [ -\omega^2 K + i \omega (M^T - M + \dot K) + (W - \dot M^T) ] = 0 . \ee
While it is a lengthy expression, it is sufficient for our purpose to consider the Minkowski limit. 
At the Minkowski limit, the dispersion relation is given by
\be \label{disp-sd}
\alpha\f{\omega^4}{M^4}  
+ \mk{ \alpha \berep_2 \f{k^2}{M^2} + \berep_0 + \tilde \berep_0 \f{M^2}{\Mpl^2} } \f{\omega^2}{M^2} 
+ \alpha \garep_4 \f{k^4}{M^4} 
+ \garep_2 \f{M^2}{\Mpl^2} \f{k^2}{M^2} =0, 
\ee
where we used the background equations~\eqref{bgeq}, and 
defined the dimensionless coefficients as
\begin{align} 
\berep_2 &= -\f{2 (f_2 + f_{2x}) }{f_2 + 3 f_{2x}}, \quad
\berep_0 = -\f{4 f_{0xx} f_2^2}{(f_2 + 3 f_{2x})^2}, \quad
\tilde \berep_0 = -\f{3 f_2  (f_{1x}^2 + \alpha f_{0xx})}{(f_2 + 3 f_{2x})^2}, \notag\\
\garep_4 &= \f{(f_2 + f_{2x})^2}{(f_2 + 3 f_{2x})^2}, \quad 
\garep_2 = -\f{f_2 (f_{1x}^2 + \alpha f_{0xx} ) }{(f_2 + 3 f_{2x})^2}  .
\end{align}
Clearly, for $\alpha=0$ we have one degree of freedom but for $\alpha\ne 0$ 
there are two branches for $\omega^2$.
Below we explicitly see that they exist at low and high energy scales, the latter of which is the apparent Ostrogradsky ghost.

Regarding $M$ as a cutoff scale, i.e.\ $\omega/M, k/M, M/\Mpl \ll 1$,  
we can solve \eqref{disp-sd} and obtain two branches of $\omega^2/M^2$:
\begin{align} \label{mode1}
\f{\omega^2}{M^2} &\simeq -\f{1}{\berep_0} \mk{ \garep_2 \f{M^2}{\Mpl^2} \f{k^2}{M^2} + \alpha \garep_4 \f{k^4}{M^4} }  \notag\\
&= - \f{(f_{1x}^2 + \alpha f_{0xx} )}{4 f_{0xx} f_2 } \f{M^2}{\Mpl^2} \f{k^2}{M^2} 
+ \f{\alpha(f_2 + f_{2x})^2 }{4 f_{0xx} f_2^2 } \f{k^4}{M^4} 
\end{align}
or 
\begin{align} \label{mode2}
\f{\omega^2}{M^2} &\simeq -\f{\berep_0}{\alpha} \notag\\
&= \f{1}{\alpha} \f{4 f_{0xx} f_2^2 }{(f_2 + f_{2x})^2 } .
\end{align}

The second mode~\eqref{mode2} does not satisfy the assumption $\omega/M\ll 1$, so it is beyond the regime of the validity of EFT.
This mode is also characterized by the fact that it diverges at the limit $\alpha= 0$.
This is actually a typical behavior of the Ostrogradsky ghost associated with nondegenerate higher-derivative term~\cite{Woodard:2015zca}.
Indeed, this mode exists only for theory with $\alpha\ne 0$, which violates the degeneracy condition.
Therefore, as expected, the scordatura DHOST model possesses the apparent Ostrogradsky ghost above the EFT cutoff scale.

Let us focus on the first mode~\eqref{mode1}.
This mode is consistent with the assumption $\omega/M\ll 1$, and hence lives at low energy.
In contrast to the high energy mode~\eqref{mode2}, this mode has a smooth limit $\alpha\to 0$ to the case of the standard DHOST.  
For $\alpha= 0$, the no-ghost condition is given by \eqref{noghost-sca}, and 
we recover the result obtained in \S\ref{sssec:1}.
Namely, if $f_{1x}=0$, the system is strongly coupled, whereas if $f_{1x}\ne 0$, the system suffers from gradient instability since $f_{0xx} f_2>0$ under the no-ghost conditions for tensor~\eqref{noghost-ten} and scalar~\eqref{noghost-sca}.
By introducing the scordatura term with $\alpha\ne 0$, for the regime $k^2/M^2\gg M^2/\Mpl^2$ the issue can be cured if $\alpha>0$. 
For $\alpha>0$, the remaining instability is IR one
analogous to the standard Jeans instability and thus is harmless (see \cite{ArkaniHamed:2003uy,ArkaniHamed:2005gu}).
The result is consistent with the EFT decoupling limit analysis in \S\ref{ssec:dslim} as well as the EFT linear perturbation theory away from the decoupling limit in Appendix~\ref{sec:lin} (see \eqref{disp-eftlin}).

\section{Discussion}
\label{sec:conc}

In the context of scalar-tensor theories we have revisited the issue of strong coupling around stealth solutions. We have pointed out an interesting role of weak and controlled violation of the degeneracy condition dubbed scordatura, which fixes the pathological dispersion relation in stealth backgrounds to healthy one and helps to make the strong coupling scale sufficiently high, while the Ostrogradsky ghost associated to the violation of degeneracy condition is adjusted to show up above the EFT cutoff scale. A scordatura DHOST theory thus realizes a ghost condensation near stealth solutions while it behaves as a usual DHOST theory away from them. 
We have illustrated the scordatura mechanism first by using the EFT action in \S\ref{sec:scs} and Appendix~\ref{sec:lin}, and then in the context of a specific class of DHOST theories in \S\ref{sec:ddh}.

As a strategy for the analysis we have employed the EFT to describe perturbations. An advantage of this is that the EFT is universal and thus makes our argument applicable to a wide class of theories. On the other hand, the symmetry assumed to derive the EFT restricts our consideration to the asymptotic region. Therefore, as a future work it is important to confirm the results of the present paper not only in the asymptotic region but also in the bulk of the geometry by using a model-dependent but more explicit methods, such as those employed in \cite{Mukohyama:2005rw,Minamitsuji:2018vuw,deRham:2019gha}.

Without the scordatura mechanism the vanishing sound speed and the strong coupling would be inevitable in a wide class of theories and stealth backgrounds, which notably includes the stealth black hole solutions in DHOST theories respecting $c_t=c$ and no graviton decay. Indeed, without the scordatura mechanism we found that the strong coupling scale is 
much lower than $M$ in the decoupling limit. 
Here, we have supposed that the action of the system in the decoupling limit is parameterized by the scale $M$ and dimensionless parameters of order unity.
Away from the decoupling limit, the strong coupling scale may become nonzero but should still be rather low since this is due to tiny corrections suppressed by negative powers of $M_{\rm Pl}^2$. Moreover, away from the decoupling limit we found that the sound speed squared receives a negative correction of order $M^2/M_{\rm Pl}^2$. The negativity is probably due to the attractive nature of gravity and thus is expected to be universal. This leads to gradient instability but can be applied to a rather narrow window below the low strong coupling scale. For this reason, we consider the lowness of the strong coupling scale more problematic than the tiny negative sound speed squared.

The consideration leading to these conclusions without the scordatura mechanism still leaves some possible loop holes (see \cite{Motohashi:2019sen,Takahashi:2019oxz,Minamitsuji:2019shy} for other type of stealth solutions to which the consideration in the present paper does not directly apply), 
one of which is to consider theories that are not consistent with gravitational wave observations or cubic order DHOST theories, and another of which is to consider stealth backgrounds with non-vanishing spatial derivative of the scalar field in the cosmic frame at infinity such as Case 2-$\Lambda$ solution.
While the former is only of theoretical interest, the latter 
may lead to some interesting signatures through statistical anisotropies of cosmological observables. However, the latter case is reduced to the case considered in this paper in the limit of vanishing spatial derivative in the cosmic frame at infinity or in the limit of vanishing cosmological constant. Hence, the strong coupling scale in the decoupling limit for the latter case in the asymptotic region should be suppressed at least by some positive powers of the remnant spatial derivative of the scalar field as well as some positive powers of the cosmological constant, meaning that the strong coupling scale in the latter case without the scordatura mechanism should also be rather low. It is nonetheless important to confirm this expectation explicitly as a future work.

In summary, without scordatura, most (if not all) of phenomenologically viable stealth solutions suffer from the strong coupling problem. This problem can be cured by the additional higher-derivative term due to the scordatura mechanism. The scordatura degenerate theory is a natural realization of EFT, and resolves the issue existing in the standard degenerate theory.

While we have demonstrated the scordatura mechanism for the spatial infinity limit of a stealth solution which can be recast to static de Sitter chart in Einstein frame in \S\ref{sec:scs}
and a specific class of DHOST theory in \S\ref{sec:ddh}, in general the introduction of higher-derivative term would affect to the strong coupling scale and the dispersion relation.
Therefore, it is natural to expect that it would work for a wider class of stealth solutions and theories such as other class of DHOST theory, or Horndeski/GLPV subclass.

An important limitation is that we have assumed the existence of timelike scalar field.
As stressed in \S\ref{ssec:tmsp}, the logic would not work if $\pa_\mu\phi$ is constant and spacelike.
Therefore, it is conceivable that the strong coupling for the stealth solution with the spacelike profile $\phi=\phi(r)$ with $X\ne {\rm const.}$ in non-shift-symmetric theory~\cite{Minamitsuji:2018vuw} could not be resolved by the scordatura mechanism.

While we adjusted the mass scale of the Ostrogradsky ghost associated with the scordatura term above the EFT cutoff scale, one may still think that the introduction of the scordatura term causes some physical process such as accretion of the scalar field, and leads to a significant difference from GR metric. 
However, within the validity of the regime of EFT, deviation from the stealth solution in the standard DHOST model should remain sufficiently small, and we expect that the evolution of the stealth black hole would also remain slow, as far as the deviation from the degeneracy condition is under controll. 
Indeed, in the case of ghost condensate, the accretion of the scalar field for the stealth solution was shown to be sufficiently slow so that it cannot be distinguished from observations~\cite{Mukohyama:2005rw}.
We expect the same scenario for the stealth solution in the scordatura degenerate theory.
Therefore, in practice, one could employ the stealth solution in standard degenerate theory as an approximation of the stealth solution in the scordatura degenerate theory.

Further, based on \cite{Mukohyama:2005rw}, it was shown in \cite{Mukohyama:2009rk,Mukohyama:2009um} that the generalized second law of black hole thermodynamics was recovered for a stealth solution in ghost condensate due to the existence of higher-derivative interaction.
It would be also intriguing to clarify if the same result holds for the stealth solution in the scordatura degenerate theory.
We leave these works for future exploration.

\acknowledgments

This work was supported in part by Japan Society for the Promotion of Science (JSPS) Grants-in-Aid for Scientific Research (KAKENHI) No.\ JP17H06359 (H.M., S.M.), No.\ JP18K13565 (H.M.), No.\ JP17H02890 (S.M.), and by World Premier International Research Center Initiative (WPI), MEXT, Japan (S.M.).

\appendix

\section{de Sitter charts}
\label{app:dS}

In general, de Sitter spacetime can be expressed in several ways, which cover whole or some part of the Penrose diagram of de Sitter spacetime.
Among the various charts,
the flat chart is given by
\be \label{flatdS} ds^2 = -dt^2 + e^{2Ht} \delta_{ij} dx^idx^j , \ee
whereas the static chart of de Sitter is given by 
\be ds^2 = - \mk{1-\f{R^2}{R_0^2}} dT^2 + \mk{1-\f{R^2}{R_0^2}}^{-1} dR^2 + R^2 d\Omega^2 ,\ee
the latter of which amounts to the spatial infinity limit of the Schwarzschild--de Sitter spacetime.

The coordinate transformation from the flat chart to the static chart is given by
\be t = T + \f{R_0}{2} \ln \mk{1-\f{R^2}{R_0^2}}, \quad r = \f{R}{ \sqrt{1 - R^2/R_0^2}} e^{-T/R_0} . \ee
Therefore, de Sitter solution with unitary gauge scalar field $\phi = qt$ in the flat chart can be transformed to de Sitter solution with the scalar field profile $\phi = qT + \psi(R)$ with $\psi(R) = \f{qR_0}{2} \ln \mk{1-\f{R^2}{R_0^2}}$, for which we have $\psi'(R) = - \f{R/R_0}{1-R^2/R_0^2}$.

Considering the limit of spatial infinity of the stealth Schwarzschild--de Sitter solution with $X=X_0={\rm const.}$, the metric approaches the static chart of de Sitter metric and the scalar field follows $\psi'(R) \to \pm \sqrt{q^2+X_0(1-\Lambda R^2/3)} / (1-\Lambda R^2/3)$.  
In particular, for Case 1-$\Lambda$ solution with $X_0=-q^2$~\cite{Motohashi:2019sen}, we have $\psi'(R) \to \pm q \sqrt{\Lambda R^2/3} / (1-\Lambda R^2/3)$.
This asymptotic behavior coincides with the one for the stealth solution obtained above by the transformation of the stealth de Sitter solution with $\phi = qt$.
Therefore, by using the inverse transformation, for the Case 1-$\Lambda$ solution in static chart can be transformed into the one with $\phi = qt$ in the flat chart.
However, it is not the case for Case 2-$\Lambda$ solution with $X_0\ne -q^2$.

\section{Linear perturbation theory of EFT action at Minkowski limit}
\label{sec:lin}

Let us investigate the EFT action~\eqref{eqn:unitary-gauge-action} without employing the decoupling limit, i.e.\ taking into account metric perturbations with the scalar-perturbed flat FLRW metric~\eqref{FLRWpert} with the gauge fixing condition $E=\delta\phi=0$.
We also take the Minkowski limit so that we can avoid the ambiguity of a choice of gauge as well as discuss stability including lower-order $k$ terms.
Since $\lambda_2$ term in \eqref{eqn:unitary-gauge-action} is third order, it does not appear in the quadratic action.
Employing the notation~\eqref{deflam},
the quadratic terms of the EFT action~\eqref{eqn:unitary-gauge-action} can be written as
\be S^{(2)} = \int d^4x \Bigg[  -\mk{ \Mpl^2 + \f{(3 \alpha + 2 \gamma) M^2}{2} } ( 3 \dot\Psi^2 + 2 k^2 B \dot\Psi ) 
+ k^2 \Mpl^2 \Psi (2 \Phi + \Psi) 
+ \beta M^3 \Phi (k^2 B + 3 \dot\Psi) 
+ \f{M^4}{2} \Phi^2 - \f{\alpha}{2} k^4 M^2 B^2 \Bigg]. \ee

Assuming $\alpha+\beta^2\ne 0$, we can solve constraint equations for $B$ and $\Phi$ to obtain 
\be \label{solbp} B = -\f{[2 \Mpl^2 + (3 \alpha + 3 \beta^2 + 2 \gamma) M^2 ] M \dot\Psi + 2 \beta k^2 \Mpl^2 \Psi}{k^2 M^3 (\alpha + \beta^2)}, \quad 
\Phi = \f{2 \beta (\Mpl^2 + \gamma M^2) M \dot\Psi 
- 2 \alpha k^2 \Mpl^2 \Psi}{M^4 (\alpha + \beta^2)} .
\ee
Plugging them back into the quadratic action, we obtain 
\be S^{(2)}= \int d^4x \kk{  
\f{\Mpl^4 (1 + \gamma \mu^2) [2 + (3 \alpha + 3 \beta^2 + 2 \gamma) \mu^2]}{ M^2 (\alpha + \beta^2) } \dot\Psi^2
+ \mk{ k^2 \Mpl^2 - \f{2\alpha}{\alpha + \beta^2}  \f{ k^4 \Mpl^4 }{M^4 }  } \Psi^2  } ,\ee
where $\mu=M/\Mpl$.
Thus, at the leading order of $M/\Mpl$, no-ghost condition is given by
\be M^2 (\alpha + \beta^2) >0, \ee
and the no gradient instability condition is given by   
\be  \f{2\alpha}{M^4(\alpha + \beta^2)}>0. \ee
The corresponding dispersion relation at the leading order of $M/\Mpl$ is given by
\be \label{disp-eftlin}
\f{\omega^2}{M^2} \simeq - \f{1}{2}(\alpha + \beta^2) \f{M^2}{\Mpl^2} \f{k^2}{M^2} + \alpha \f{k^4}{M^4} ,
\ee
This corresponds to $H=0$ limit of the dispersion relation \eqref{disp-gen} at the decoupling limit together with the correction of $\mathcal{O}(M^2/\Mpl^2)$ to sound speed squared away from the decoupling limit.

If $\alpha \f{k^2}{M^2}\ll |\alpha + \beta^2| \f{M^2}{\Mpl^2}$, notably including the case $\alpha=0$, we have 
\be \label{disp-lin1} \f{\omega^2}{M^2} \simeq - \f{1}{2}(\alpha + \beta^2) \f{M^2}{\Mpl^2} \f{k^2}{M^2} .  \ee
As a very rough estimation of $E_{\rm cubic}$, let us employ the estimation \eqref{Ecubic} at the decoupling limit, while we are addressing the case away from the decoupling limit.
We then obtain 
\be E_{\rm cubic} \sim |\alpha + \beta^2|^{7/8}  \mk{\f{M}{\Mpl} }^{7/4} M \ll M . \ee
Therefore, the strong coupling scale is expected to be much lower than $M$, and its supression factor is about $\sim M^2/\Mpl^2$.
As a special case, for $\alpha=0$ and $\beta\ne 0$, the system exhibits gradient instability in the rather narrow window below the low strong coupling scale $E<E_{\rm cubic}$.
In this regime, however, the timescale of the gradient instability is very long $E^{-1}\gg M^{-1}$ so the instability is rather weak.

In contrast, if $|\alpha + \beta^2| \f{M^2}{\Mpl^2}\ll \alpha \f{k^2}{M^2}$, requiring $\alpha\ne 0$, we have 
\be \label{disp-lin2} \f{\omega^2}{M^2} \simeq \alpha \f{k^4}{M^4} , \ee
which matches the Minkowski limit of the dispersion relation~\eqref{disp-3} in the decoupling limit.
Therefore, assuming $\alpha=\mathcal{O}(1)$, the decoupling limit corresponds to $k^2/M^2 \gg M^2/\Mpl^2$.
Similar to the previous case, a very rough estimation employing \eqref{Ecubic} yields
\be E_{\rm cubic} \sim \alpha^{7/2} M , \ee
which can be as high as $M$ provided that $\alpha=\mathcal{O}(1)$.

In summary, away from the decoupling limit, without the scordatura term, the strong coupling scale is rather low and the system exhibits gradient instability in the rather narrow window below this low strong coupling scale.
This issue can be cured if we employ the scordatura term with $\alpha>0$.
If $\alpha=\mathcal{O}(1)$, the decoupling limit is $k^2/M^2 \gg M^2/\Mpl^2$ for which \eqref{disp-3} is recovered.
The result we obtained here is consistent with the decoupling limit analysis in \S\ref{ssec:dslim} as well as the stealth solution in the scordatura DHOST theory in \S\ref{sec:pert}.

Finally, for the remaining exceptional case $\alpha+\beta^2= 0$, the two constraint equations are degenerate, and hence one cannot solve $B$ and $\Phi$ at the same time, unlike \eqref{solbp}.
Integrating out $\Phi$ and performing integration by parts, we obtain
\be S^{(2)} = \int d^4x \kk{ -3 (\Mpl^2 + \gamma M^2) \dot\Psi^2 + 
k^2 \mk{ \Mpl^2 - 2 k^2 \f{\Mpl^4}{M^4} } \Psi^2 - 
2 k^2 B \mk{ (\Mpl^2 + \gamma M^2) \dot\Psi + \beta k^2 \f{\Mpl^2}{M}  \Psi }  } .  \ee 
Along the same line as \eqref{tL}, 
we can replace $\dot\Psi$ to an auxiliary field $Q$ by adding a new term $\lambda(Q-\dot\Psi)$ to the Lagrangian, where $\lambda$ is a Lagrange multiplier.
Integrating by parts to move the time derivative from $\Psi$ to $\lambda$, the quadratic action can be written down in terms of $Q,\Psi,B, \lambda, \dot\lambda$.
Now we can solve constraint equations for $Q,\Psi,B$ and express them in terms of $\lambda, \dot\lambda$.
Plugging them back into the quadratic action and performing integration by parts, we arrive at
\be S^{(2)} = \int d^4x \f{M^2 \mu^2 (1 + \gamma \mu^2)^2 \dot\lambda^2 + k^4 \beta^2 \mu^2 \lambda^2 }{4 k^2  (1 + \gamma \mu^2) [\Mpl^2 k^2 \{2 + (3 \beta^2 + 2 \gamma) \mu^2\} - M^4 (1 + \gamma \mu^2)]} , \ee 
and at the leading order of $M/\Mpl$ no-ghost condition is $M^2>0$.
The corresponding dispersion relation at the leading order of $M/\Mpl$ is given by
\be \label{disp-lin3} \f{\omega^2}{M^2} \simeq -\beta^2 \f{k^4}{M^4} . \ee
Therefore, while the reduction of the quadratic action for the case $\alpha+\beta^2= 0$ requires a different treatment from the one for the case $\alpha+\beta^2\ne 0$ above, the outcome \eqref{disp-lin3} is consistent with \eqref{disp-lin2} in the limit $\alpha+\beta^2= 0$.
However, in the present case, the system suffers from either gradient instability or strong coupling, and hence the scordatura with $\alpha=-\beta^2$ does not solve the issue.

\bibliographystyle{apsrev4-1}
\bibliography{ref-BH}

\end{document}